# Baseline design of the filters for the LAD detector on board LOFT

M. Barbera[*a,b], B. Winter[c], J. Coker[c], M. Feroci[d,e], T. Kennedy[c], D. Walton[c], S. Zane[c]

[a*]Università degli Studi di Palermo, Dipartimento di Fisica e Chimica, Via Archirafi 36, 90123 Palermo, Italy; [b]Istituto Nazionale di Astrofisica, Osservatorio Astronomico di Palermo, Piazza del Parlamento 1, 90134 Palermo, Italy; [c]Mullard Space Science Laboratory, University College London, Holmbury St. Mary, Dorking, Surrey RH5 6NT, United Kingdom; [d]Istituto Nazionale di Astrofisica, Istituto di Astrofisica e Planetologia Spaziale Roma, Via del Fosso del Cavaliere 100, 00133 Roma, Italy; [e]Istituto Nazionale di Fisica Nucleare, Sezione di Roma Tor Vergata, Via della Ricerca Scientifica 1 - 00133 Roma, Italy

## ABSTRACT

The Large Observatory for X-ray Timing (LOFT) was one of the M3 missions selected for the phase A study in the ESA's Cosmic Vision program. LOFT is designed to perform high-time-resolution X-ray observations of black holes and neutron stars. The main instrument on the LOFT payload is the Large Area Detector (LAD), a collimated experiment with a nominal effective area of ~10 m$^2$ @ 8 keV, and a spectral resolution of ~240 eV in the energy band 2-30 keV. These performances are achieved covering a large collecting area with more than 2000 large-area Silicon Drift Detectors (SDDs) each one coupled to a collimator based on lead-glass micro-channel plates.
In order to reduce the thermal load onto the detectors, which are open to Sky, and to protect them from out of band radiation, optical-thermal filter will be mounted in front of the SDDs. Different options have been considered for the LAD filters for best compromise between high quantum efficiency and high mechanical robustness. We present the baseline design of the optical-thermal filters, show the nominal performances, and present preliminary test results performed during the phase A study.

**Keywords:** X-ray, mission, LOFT, LAD, silicon drift detector, optical blocking filters

## 1. INTRODUCTION

The Large Observatory for X-ray Timing (LOFT) [1][2] was one of the candidate missions for the M3 launch opportunity in ESA's Cosmic Vision program selected for the phase A study. LOFT, designed to perform high-time-resolution X-ray observations of compact sources, brings on-board two instruments:

1. Large Area Detector (LAD) [3], a collimated experiment with a nominal effective area of ~10 m$^2$ @ 8 keV, and a spectral resolution of ~240 eV in the energy band 2-30 keV;
2. Wide Field Monitor (WFM) [4], a coded mask camera with solid state-class energy resolution, which observes a large fraction of the sky to prompt changes in the state of targets of interest, or to discover and localize new sources. If a relevant transient is detected, the viewing direction of LOFT can be modified to observe this source with the LAD.

The gain factor larger than 20 on the LAD effective area, with respect to any previous or currently investigated X-ray timing missions, relies on a modular approach, from payload large deployable panels to individual detector units based on the innovative technology of large-area Silicon Drift Detectors (SDDs) designed on the heritage of the ALICE experiment at CERN/LHC [5][6]. Figure 1 is an exploded view of the modular LAD concept. 16 detector tales 11 x 7 cm$^2$ are assembled on a single module, and about 120 modules are needed to obtain the total effective area.

---

[*] marco.barbera@unipa.it

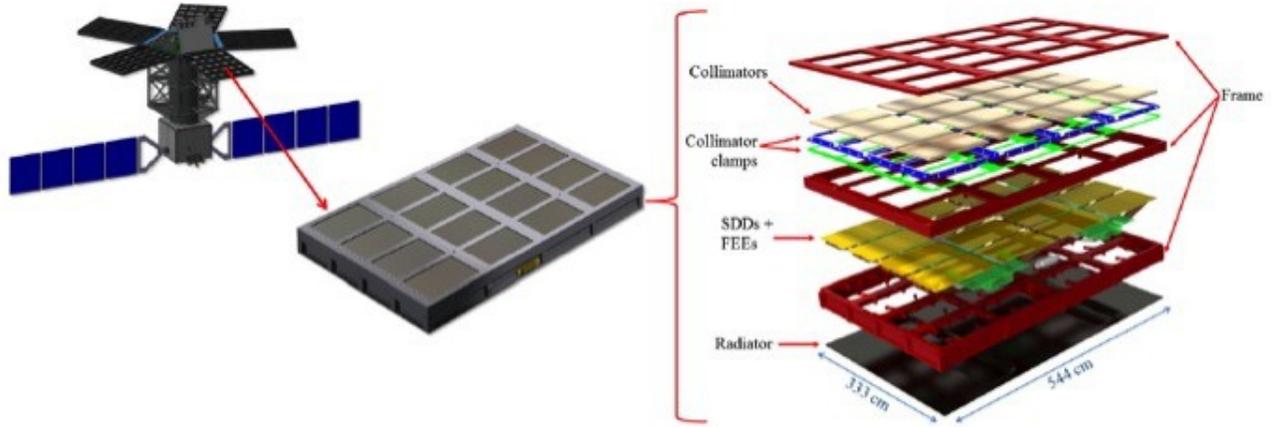

**Figure 1.** LAD modular design. Each deployable panel (left) supports ~ 25 modules (middle). Each module (right) holds 16 detectors together with the front-end electronics, module back-end electronics, power supply unit, and the relevant supporting structure and shielding.

The SDDs are able to identify the time of arrival of an X-ray photon with a time accuracy better than 10 μs and an energy resolution of 240 eV FWHM. Coupling such detectors to an X-ray collimator allows also to narrow the field of view below 1° FWHM, large enough to allow for pointing uncertainties, and small enough to reduce the aperture background (cosmic diffuse X-ray background) and the risk of source confusion. The collimator is based on micro-channel plates similar to those developed for Bepicolombo[7], consisting of a 5 mm thick sheet of Lead glass with a large number of square pores, ~83 **μ**m pore width and ~16 **μ**m wall thickness, giving an open area ratio of 70%. The stopping power of lead in the glass over the large number of walls that off-axis photons need to cross is effective in collimating X-rays below 30 keV.

The adopted LAD design based on SDD detectors and MCP collimators has another significant advantage in the low weight which is < 10 Kg/m$^2$ of effective area, nearly a factor 10 lower than other experiments like RXTE/PCA[8].

## 2. OPTICAL-THERMAL BLOCKING FILTERS

The LAD SDDs need to be operated at temperatures lower than -10 C° to minimize the leakage current and obtain optimal performances over the lifetime of the experiment. An extensive design trade off was performed to assess optimal configurations for the module from a thermal point of view [9]. A radiator is located at the back of each module, and both the radiator and the module box are covered with multilayer mirror tape to optimize heat rejection combined with minimizing solar heat absorption.

Radiation heat load has also to be blocked on the detector side of the module. The radiation load from hot surfaces in the temperatures range below 300 K is mainly in the IR region between 1 μm and few cm. A layer of 10 nm of aluminum provides already a reflectivity larger than 95% in this IR region, and the reflectivity marginally changes by increasing the thicknesses above 20 nm. Aluminum is, however, oxidized and as a first approximation the oxide layer becomes nearly transparent. In the past, we have estimated an oxidation layer of the order of 50 Å on each side of the aluminum layer both by UV/Vis transmission measurements, and X-ray photoelectron spectroscopy measurements conducted on samples of the XMM-Newton EPIC camera thin and medium filters [10][11].

An aluminum layer of 40 nm will be directly deposited onto the MCP collimator to behave as a thermal shield on the detector side. Figure 2 shows the reflection (left panel) and transmission of a layer of aluminum 30 nm thick representative of the expected performance of a 40 nm layer of aluminum oxidized on both sides. The calculations are derived by the use of the matrix formulation of the boundary conditions of the electromagnetic field [12], with the refractive index of aluminum derived from three different sources for the wavelength range $0.01653 \div 32$ μm[13], $33.333 \div 200$ μm [14], and $248 \div 1240$ μm [15], respectively.

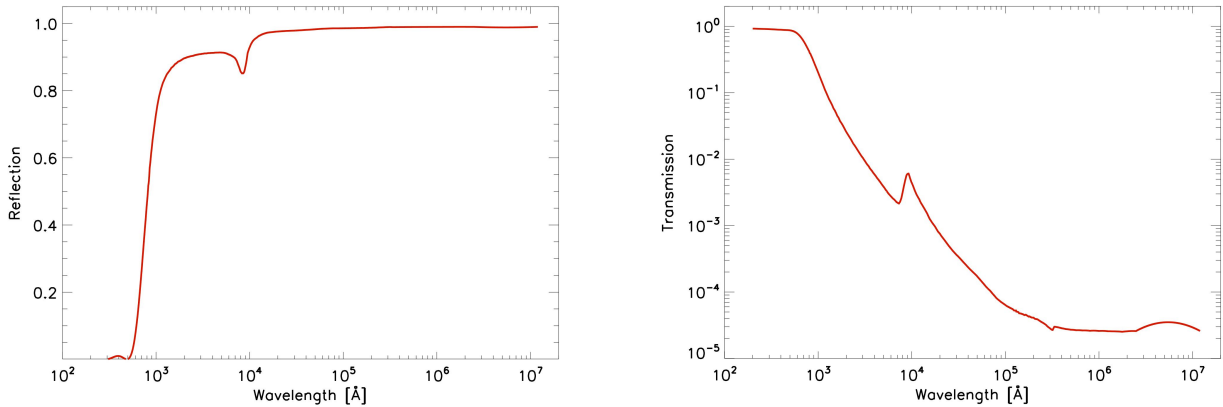

**Figure 2.** Calculated reflection of a single layer of aluminum 30 nm thick (left panel), and calculated transmission for the same film of aluminum (right panel).

While aluminum provides good IR rejection it is not sufficient to protect the SDDs from Vis/UV radiation from pointed sources or aperture background which can introduce a detector energy resolution degradation or increase the background. For this reason, a thin (~1 μm) polyimide film coated with 40 nm Al has been included in the LAD design. Such filter is similar as the filter flying on Chandra to protect the HRC-I from UV and low energy particles [16][17], and thus as a first option we have considered to mount a free standing filter between the collimator tray and the detector tray. Given the large size of the filters we have also decided to investigate two other options providing a mechanical support to the film, either by mounting the filter on a metal mesh or mounting the filter directly onto the MCP collimator. In the last option the polyimide film has been reduced to 0.5 μm. The following list summarizes the three investigated filters.

   **option 1**:  1.0 μm polyimide + 40 nm aluminum unsupported
   **option 2**:  1.0 μm polyimide + 40 nm aluminum supported by a mesh
   **option 3**:  0.5 μm polyimide + 40 nm aluminum supported by the MCP collimator

Polyimide is quite transparent in the visible and near IR while it strongly absorbs radiation in the UV at wavelengths shorter than about 3500 Å [16]. In the far IR molecular absorption bands occur at wavelengths larger than approximately 20.000 Å that may significantly reduce the transmission[18][19]. For our purposes of designing an optical-thermal blocking filter we can conservatively consider polyimide to be nearly transparent at wavelength larger than 20.000 Å.

Figure 3 shows the total transmission in the UV/VIS/IR for the three investigated configurations, including also the the 40 nm aluminum layer deposited on the MCP. The calculated transmission curves are based on the refractive index of polyimide derived by Cavadi et al. 2001 [20] based on transmission measurements performed on Hitachi PIQL-100 polyimide samples provided by LUXEL corp., which is the same material used for the filters of the Chandra HRC [16].

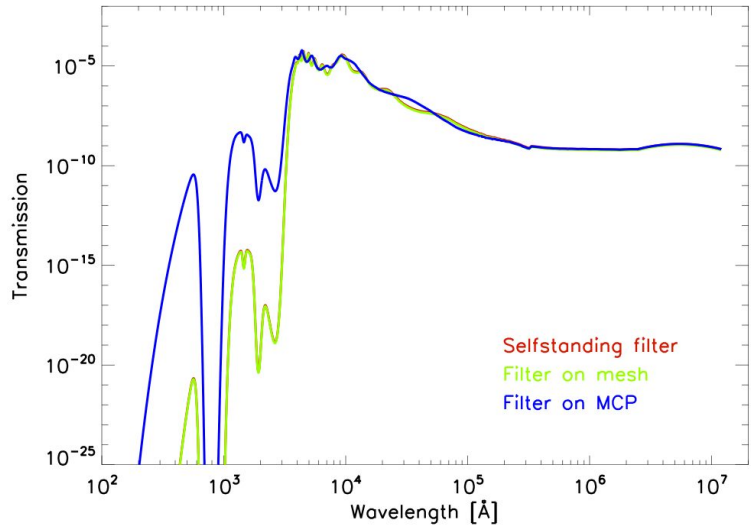

**Figure 3.** Total transmission in the UV/VIS/IR for the three investigated configurations of optical-thermal filter. The calculated transmission includes the 40 nm aluminum layer directly deposited onto the MCP collimator.

Figure 4 shows a comparison of X-ray transmission, in the low energy range of the sensitivity of LAD, for the three investigated optical blocking filter configurations. Again the transmission curves include the 40 nm aluminum layer directly deposited onto the MCP collimator.

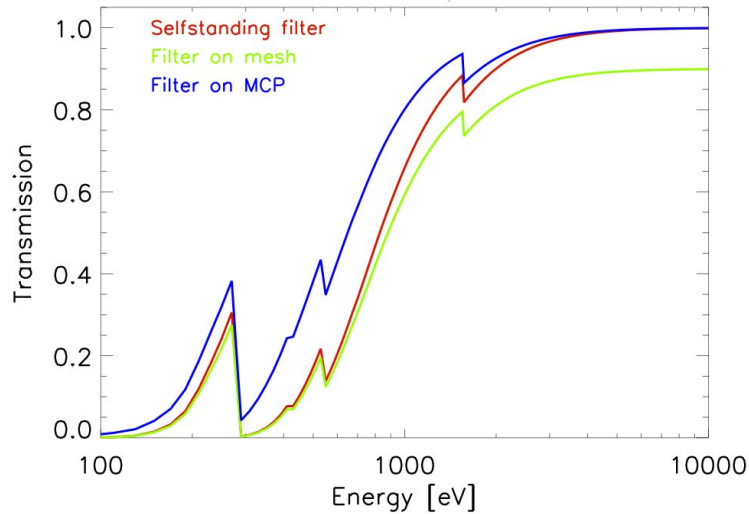

**Figure 4.** Total X-ray transmission, in the low energy range of the sensitivity of LAD, for the three investigated optical-thermal blocking filter configurations, including the 40 nm aluminum layer deposited on the MCP collimator.

The filter supported by the MCP collimator has a thinner polyimide (0.5 μm) layer which provides a lower UV rejection with respect to the unsupported and mesh supported filters. However, an attenuation of $10^6$ in the UV/Vis is usually considered sufficient for most Astrophysical sources to prevent out of band radiation to dominate the X-ray source emission [21]. For this reason, all three options can be considered safe as optical blocking filters.

The filter supported by the MCP collimator, having less polyimide is more transparent below 1 keV while at energies larger than 1 keV the filter with mesh is less efficient and would cause a 10% loss in the LAD effective area with respect to the two other options. The best X-ray performance is provided by the filter mounted directly on the MCP, however, this filter can present some more difficulties in the mechanical design and integration procedure.

# 3. FILTER FRAME DESIGN AND SAMPLE PROCUREMENT

Two test samples of the investigated filters option 1 (unsupported) and 2 (supported on mesh) have been built for the LOFT consortium by LUXEL corp. The frame design is based on that one developed for the large area UV/Ion shield (10 x 10 cm$^2$) of the CHANDRA High Resolution Camera imaging detector (HRC-I). Figure 5 shows the CAD drawing of the filter frame designed by LUXEL.

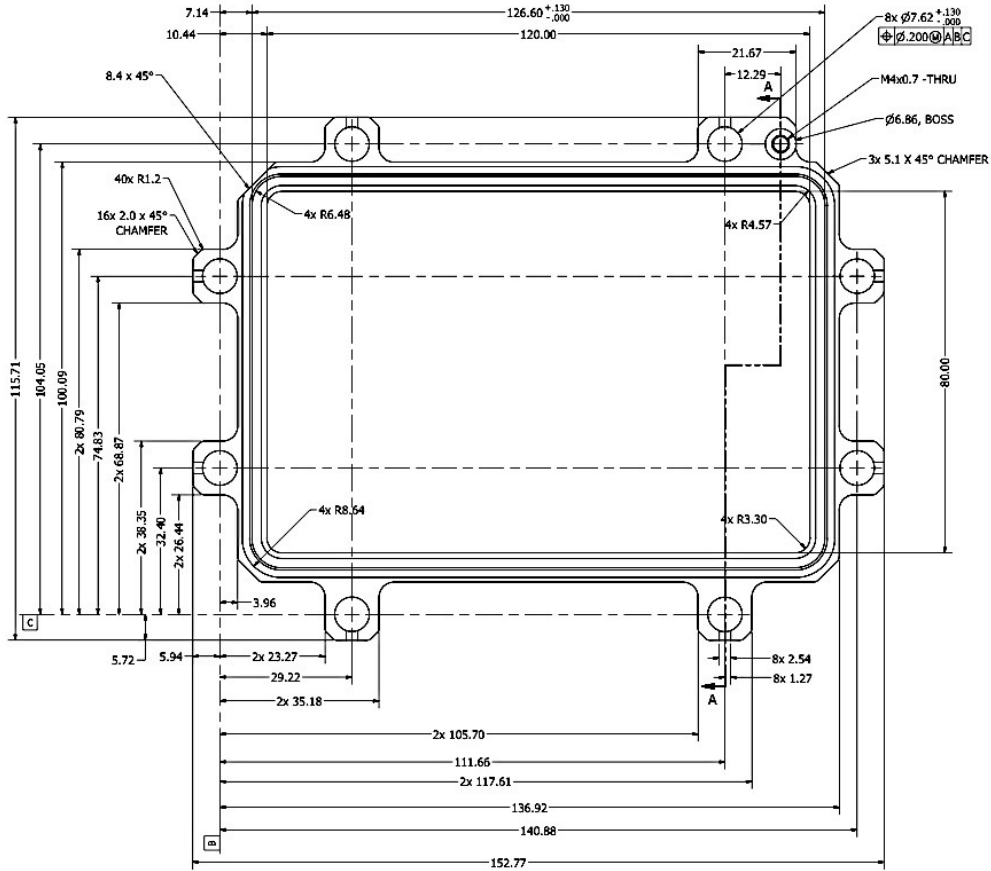

**Figure 5.** LAD filter frame design. The frame built with 304 stainless steel is based on the design developed for the large area (10 x 10 cm$^2$) unsupported filter of the Chandra HRC-I detector.

The adopted mesh for the supported filter is stainless steel with 90% open area and is the same type used on the ASTRO-H Thermal Shields[22]. Aluminized polyimide films are mounted on the filter frame according to the schematic shown in Figure 6. The mesh is on the polyimide side and the aluminum is facing the detector.

Three test samples of each one of the two filter types, previously identified as option 1 and 2, have been built by LUXEL corp. for the LOFT study phase A, namely:

1.0 μm  LUXfilm ® polyimide  + 40 nm aluminum unsupported
1.0 μm  LUXfilm ® polyimide  + 40 nm aluminum supported by a stainless steel mesh 90% open area

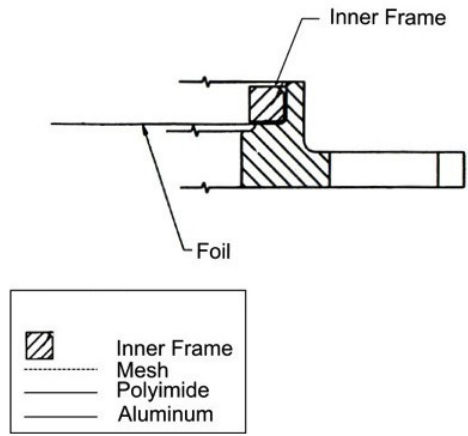

**Figure 6**. Drawing of the scheme adopted to mount the filter on the frame.

Filters of the type option 3 have not been manufactured since MCP of the needed size were not available at the time of filter procurement during the phase A study. Figure 7 shows a picture of one of the filter samples procured for testing during the LOFT phase A study.

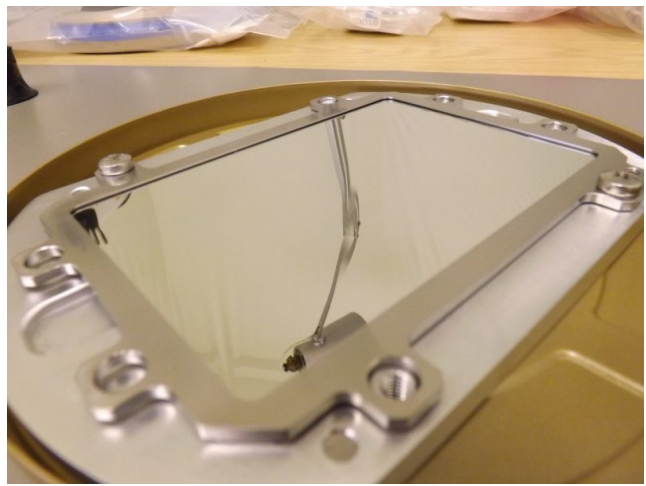

**Figure 7.** One of the LAD filter samples in its frame mounted inside its transport container

## 4. ACOUSTIC TESTS

A concern regarding mounting a thin filter in front of the SDD is whether it is capable of surviving the acoustic pressure during launch. During the launch of the spacecraft the modules are subjected to both high sound pressure levels from the main rocket engines during lift off as well as rapid depressurization of the fairing after takeoff. These two events are different in nature. The acoustic sound pressure during lift off is diffuse and random in nature, with peak frequencies centered around 250 Hz. The second event, the depressurization of the fairing, is more quasi-static and simply depends primarily on the altitude of the launch vehicle before reaching the vacuum of space. This pressure rate off change is on average 1 kPa/sec but will reach 2 kPa/Sec around 50 seconds into the launch.

In order to show that the proposed X-ray filter for the LAD would be able to survive at least the diffuse acoustic sound levels we subjected two different filters to levels up to 143 dB Overall Sound Pressure Level (OSPL). Both the two types of previously described filters were tested, namely: the one with substrate (1 μm thick LUXfilm ® polyimide coated with 40 nm of aluminum) unsupported (option 1), and the one with the same substrate supported by a stainless steel mesh with 90% open area (option 2). Each filter was mounted on the frame previously described, and the frame was then mounted inside a cavity representing the Micro-channel Pore Optics (MPO) frame and box. The exposed area within the mounting frame is 80 by 120 mm². Acoustic tests have been conducted at the Institute of Sound and Vibration Research, University of Southampton, (Highfield, Southampton, SO17 1BJ, UK)

For the LAD module one of the configurations that was looked into is to have the filter mounted as a separate item instead of it being an integral part of the MPO tile. The two different design solutions have their pros and cons regarding assembly and thermal aspects. Analysis has shown that both solutions can work from a thermal point of view. However the idea of mounting a thin filter inside a box without it being bonded onto a surface, like the MPO tile, is risky since it will be subjected to the sound pressure variations which are huge and the filter is without the backing of anything solid. A proven approach to this problem is to house the filter inside a separate enclosure, preferably under vacuum during launch, and then open the enclosure after launch. This is however impossible in the case of the LAD module for several reasons. There is no envelope inside the module to provide for a separate enclosure surrounding each filter. If one would increase the size of the module to make this possible, which in theory could be done, the increase of the overall mass would be prohibitive. The available mass simply doesn't allow for such a solution. In addition to this, having to provide for more than 2000 of these enclosures (as a worst case) is financially not remotely viable. So if the filter in front of the SDD has to be flown as a separate part, mounted on its frame, it has to be able to survive the acoustic qualification load on its own.

The best way to approach this problem is to mount the filter in such a way that it is very close to a solid surface. Pressure variations near the surface of the filter will be absorbed by the solid surface and the acoustic loading of the filter itself will be minimal. For electrical reasons, the SDD is a high voltage device and the filter is electrically conductive, the filter cannot be mounted near the surface of the SDD. This leaves the surface of the MPO itself as the only available alternative. However the surface of the MPO is not solid from a mechanical point of view. Acoustically however the pore size of less than 100 micron will effectively be solid for sound waves with a frequency below 10 kHz. Static pressure will propagate, but pressure variations in the 250 Hz range will be attenuated significantly.

Taking into account manufacture and assembly considerations the baseline for the LAD module is to mount the filters within 0.5 mm from the MPO tile. One could try and mount the filter closer to the MPO surface, but the 0.5 mm with a 0.1 mm tolerance allows for a more cost effective design, which is important when the design has to be implemented more than 2000 times.

Since the actual configuration of the module is difficult to analyze from a vibrational-acoustic point of view it was decided to build a small test model and test the idea of mounting the filter close to the MPO tile. A test model was built and the two different types of filters were tested. The enclosure had a transparent window making it easy to quickly check if the filter had survived a test run or not. Only a significant tear in the filter could be observed this way, if it appeared, but one would expect that the failure mode of the filter would involve significant tears along the interface with the mounting frame.

No representative MPO tiles were available for this test, so a dummy tile was manufactured. The dummy MPO consisted of a 5 mm thick aluminum plate with 1 mm holes. The holes are an order of a magnitude larger than the proposed MPO tile, this was accepted as a worst case. The 1 mm holes would still not be transparent, as waveguides, to sound pressure waves below 10 kHz. The filter in the test setup was mounted 0.75 mm away from this surface, again as a worst case approach. In the acoustic test facility the enclosure was suspended in the corner of the echoic chamber with three control microphones around it. The acceleration of the dummy MPO was measured in the longitudinal direction (X) and in the out of plane direction (Z). The transverse direction was measured as well (Y) however this is the direction in which the enclosure was suspended and the response is a combination of the tension in the horizontal cable and the enclosure itself and they were left out. For the filter the movement of the dummy MPO out of plane (Z) is the most important one, since that puts pressure onto the filter, along with the pressure propagating through the clearances in the enclosure. The enclosure was not hermetically sealed and sound waves could enter via 0.5 mm wide gaps between the transparent window and the aluminum back as well as between the two aluminum parts, making up the enclosure itself. Figure 8 shows some pictures of the apparatus used to conduct the acoustic tests on the two LAD filter samples.

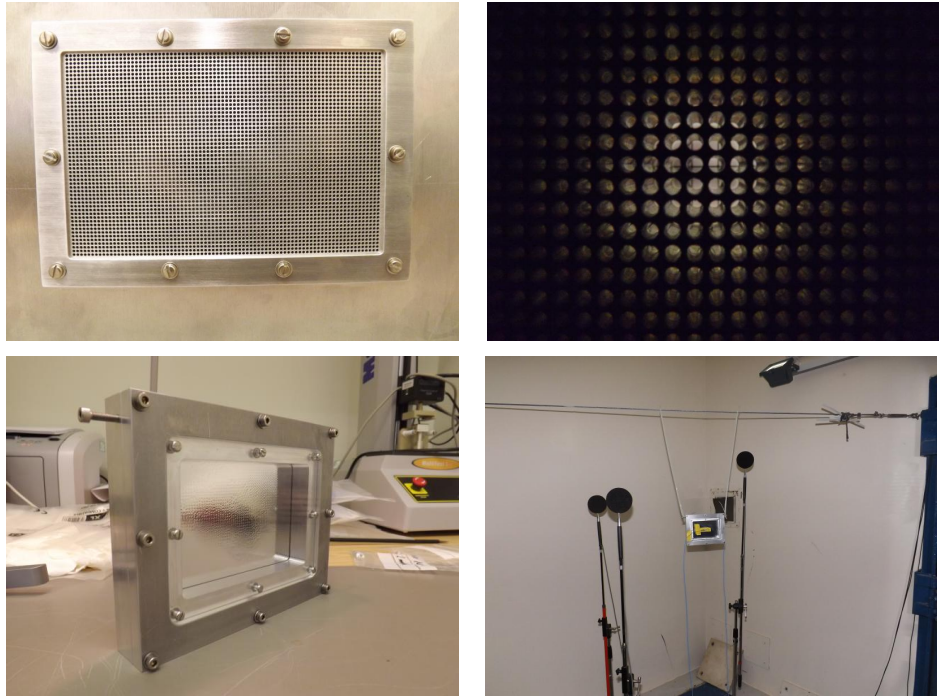

**Figure 8.** Pictures of the experimental apparatus used to perform acoustic tests of the LAD filter samples. Front view of the test enclosure with the dummy MPO in place (Top left). Detailed view of the MPO with the mesh supported filter behind it (Top right). Rear view of the enclosure with the transparent window (bottom left). The enclosure suspended inside the acoustic test facility (bottom right).

Several tests were performed on the unsupported filter (in the following Filter 1) and on the filter with a thin stainless steel mesh as additional support (Filter 2). Test run 1 involved Filter 2 with the stainless steel mesh. For this test the gap was kept at a minimum 0.75 between dummy MPO and the filter. The dummy MPO was covered in a single layer of tape to close the holes in the MPO tile with a membrane. This additional test was introduced to increase the acoustic pressure on the dummy MPO itself. Test run 2 and higher involved the bare filter, without the stainless steel mesh support. For this filter the gap distance was increased as well as the tape on the MPO removed. Table 1 lists the measured acceleration at the center of the dummy MPO. After mounting a different filter the amplification of the dummy MPO had changed. The first eigen frequency remained the same, 950 Hz, only the amplification changed. We think this is due to the clamping of the dummy settling after the first test. The amplification for the 3rd tests didn't really change, only a slight difference in amplification. The 4th test with the larger gap and the tape removed showed a significant drop in amplification of the dummy MPO. In table 1, X is pointing along the longitudinal line of the dummy MPO, perpendicular to the suspension of the enclosure.

Table 1. the response of the dummy MPO for different configurations

| Test run | | Test Level | Duration seconds | Gap mm | MPO taped | X g-rms | Z g-rms |
|---|---|---|---|---|---|---|---|
| 1 | Filter 2 | 144 dB | 120 | 0.75 | yes | 1.9 | 18.8 |
| 2 | Filter 1 | 144 dB | 120 | 0.75 | yes | 2.4 | 9.0 |
| 3 | Filter 1 | 144 dB | 60 | 1.45 | yes | 2.4 | 8.2 |
| 4 | Filter 1 | 144 dB | 60 | 1.45 | no | 0.3 | 2.2 |

Figure 9 shows the measured acceleration at the center of the dummy MPO in test run 2.

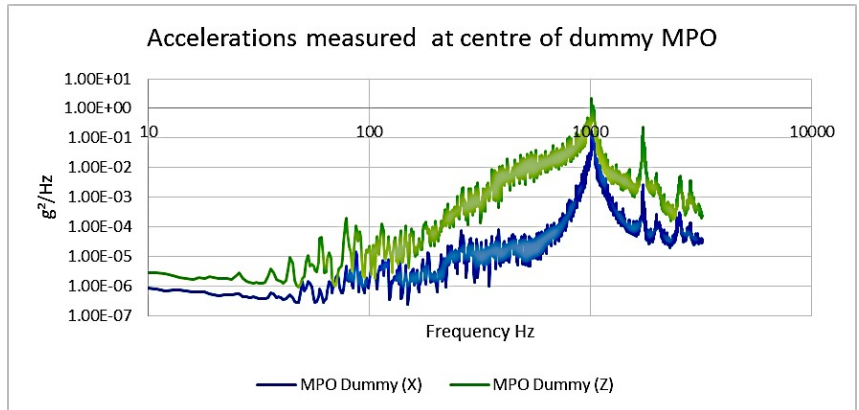

Figure 9: Response at the center of the MPO as measured during test run 2

Table 2 lists the measured sound pressure levels in dB for the different frequency bands as well as the overall sound pressure level for a typical test run. The sound pressure levels are based on the ones specified for the Soyuz launch vehicle with the ST-type fairing with 3 dB added to each band as qualification margin.

Table 2. Typical example of applied sound pressure levels during the test, controlled at 144 dB

| Octave frequency (Hz) | Sound Pressure Level (dB re 20mPa) | | | | | |
|---|---|---|---|---|---|---|
| | Microphone | | | Average SPL | Tolerance Limits | |
| | 1 | 2 | 3 | | Lower | Upper |
| 31.5 | 127.9 | 127.7 | 127.8 | 127.8 | 127.0 | 130.0 |
| 63 | 133.5 | 135.2 | 134.4 | 134.4 | 134.0 | 137.0 |
| 125 | 136.9 | 135.7 | 138.2 | 137.1 | 136.0 | 139.0 |
| 250 | 138.1 | 137.8 | 140.6 | 139.0 | 138.0 | 141.0 |
| 500 | 137.0 | 136.6 | 138.7 | 137.5 | 136.0 | 139.0 |
| 1000 | 128.7 | 129.2 | 131.2 | 129.8 | 127.0 | 130.0 |
| 2000 | 120.9 | 120.3 | 122.6 | 121.4 | 123.0 | 126.0 |
| 4000 | 114.0 | 113.7 | 115.5 | 114.5 | N/A | N/A |
| 8000 | 112.8 | 112.2 | 113.1 | 112.7 | N/A | N/A |
| OASPL | 143.1 | 142.9 | 144.8 | 143.7 | 143.0 | 146.0 |

Both filters survived all test runs. Therefore, a separate filter mounted alongside the (dummy) MPO can survive the acoustic test levels. All dimensions of the filters and local enclosure around the filter were within 10% of the current baseline design of the LAD module, the tested filter was actually larger than what we would use for the LAD. The two different types of filters both survived the acoustic test. We tested the filters with varying gap sizes (0.75 and 1.45 mm) between the filter and the dummy MPO tile. We tested with a membrane on the MPO tile as well as without, it appears that the test without the tape shows less response of the dummy MPO itself.

The test was not 100% representative of a real LAD module as the structural frequencies of the enclosure were not close to the proposed module design. But as a proof of concept and a de-risking test this activity proved to be very successful.

Clearly a thin, fragile, filter can survive the harsh acoustic test environment as imposed by many space projects when mounted close to a solid surface.

## 5. SUMMARY AND CONCLUSIONS

The Silicon Drift Detectors of the LAD experiment on-board LOFT need to be operated at temperatures < 10 C$^o$, for this reason thet need to be protected by radiation load from the spacecraft and the solar scattered light. A layer of aluminum 40 nm thick is deposited, for this purpose, directly onto the micro-channel plate collimator. The semiconductor detectors are also sensitive to UV/Vis light that also need to be blocked to prevent energy resolution degradation or loss of sensitivity to X-ray sources. Aluminum is able to block IR radiation but it is not sufficient to block UV/VIS light. A filter made of a polyimide film coated with aluminum will be mounted in between the MCP collimator and the SDD. Three different options have been considered for this filter, namely:

> **option 1**: 1.0 μm polyimide + 40 nm aluminum unsupported
> **option 2**: 1.0 μm polyimide + 40 nm aluminum supported by a stainless steel mesh 90% open area
> **option 3**: 0.5 μm polyimide + 40 nm aluminum supported by the MCP collimator

The filter supported by the MCP collimator (option 3) has a thinner polyimide layer (0.5 μm) which provides a lower UV rejection with respect to the unsupported and mesh supported filters whose polyimide film is 1 μm thick. However, since an attenuation of $10^6$ in the UV/Vis should be sufficient for most Astrophysical sources to prevent out of band radiation to dominate the X-ray source emission, all three options can be considered safe as optical blocking filters. Detailed analysis should be performed for particular sources with very low fx/fv ratio such as massive O stars, planets, and QSO's.

The filter supported by the MCP collimator, having less polyimide is more transparent below 1 keV while the filter with mesh is always less efficient and would cause a 10% loss in the LAD effective area with respect to the two other options. The best X-ray performance is provided by the filter mounted directly on the MCP, however, this filter can present some more difficulties in the mechanical design and integration procedure. The unsupported filter is a good compromise from all points of view, while filter supported by the mesh has to be considered only as a back-up solution since it significantly reduces the LAD effective area.

In order to show that the proposed X-ray filter for the LAD would be able to survive the high sound pressure levels from the main rocket engines during lift off, we subjected two different filters to levels up to 143 dB Overall Sound Pressure Level (OSPL). Two types of filters previously described were tested, namely; the one with substrate (1 μm thick LUXfilm ® polyimide coated with 40 nm of aluminum) unsupported (option 1), and the one with the same substrate supported by a stainless steel mesh with 90% open area (option 2). Each filter was mounted in its frame, and the frame was then mounted inside a cavity representing the Micro-channel Plate (MPC) collimator frame and box. The exposed area within the mounting frame is 80 by 120 mm².

Several tests were performed with different filters, different distances between the filter and the dummy MCP collimator, with and without a tape to close the MCP micro pores. Both filters survived all the acoustic tests. The tests were not 100% representative of a real LAD module as the structural frequencies of the enclosure were not close to the proposed module design. But as a proof of concept and a de-risking test this activity proved to be very successful. Clearly a thin, fragile, filter can survive the harsh acoustic test environment as imposed by many space projects when mounted close to a solid surface.

## ACKNOWLEDGEMENTS

We acknowledge fruitful suggestions and advice by prof. George Fraser, a dear friend and colleague, and dedicate this work to his memory. We thank LUXEL corp. for supporting the LAD filter design activity. We acknowledge partial support by the Italian Space Agency (under contract I/021/12/0-186/12), United Kingdom Space Agency, INAF, and INFN.